\documentclass[aps,prb,showpacs,twocolumn]{revtex4-1}
\usepackage[utf8x]{inputenc}
\usepackage{amssymb}
\usepackage{graphicx}
\begin{document}
\title{Muon Spin Rotation in Pauli Limited Superconductors}

\author{V. P. Michal, A. Yaouanc and P. Dalmas de R\'eotier}
\affiliation{Commissariat \`a l'Energie Atomique,
INAC/SPSMS, 38054 Grenoble, France}
\date{\today}

\begin{abstract}
The formalism for analysing the magnetic field distribution in Pauli limited superconductors developed earlier is applied to the field dependence of the vortex lattice static linewidth measured in Muon Spin Rotation ($\mu$SR) experiments. In addition to writing analytical formulae for the static linewidth for the vortex structure in the limit of independent vortices (i.e. moderate magnetic fields), we use Abrikosov's analysis to describe the field variations of the static linewidth at the approach of the superconductor to metal transition in the limit where the critical field is determined by Pauli depairing. 
\end{abstract}
\maketitle

It has been proposed in a Letter by Spehling et al. \cite{Spehling} that anomalous variations of the static linewidth $\sigma_s$ measured by Muon Spin Rotation ($\mu$SR) on the heavy fermion superconductor CeCoIn$_5$ is due to coupling between superconducting and antiferromagnetic order. The static linewidth was found to increase with applied field in a wide range below the upper critical field $H_{c2}$ whereas the theoretical description available so far \cite{Yaouanc} predicted a monotoneous decrease in $\sigma_s$. Here it is shown that the chief effect responsible for such a behavior is a novel mechanism that was analyzed before\cite{Michal} and observed in Small Angle Neutron Scattering (SANS) experiments \cite{Bianchi,White} on the Vortex Lattice (VL) of CeCoIn$_5$ at low temperature and high magnetic field applied along the c-axis. Prior publication of analytical expressions resulting from the Ginzburg-Landau formulation\cite{Michal}, numerical results\cite{Machida} based on the Bogoliubov equations were reported. 

Because of a large electron effective mass ($m^\ast\simeq100m_e$), the diamagnetic screening supercurrents generated by the electron Zeeman spin response under a field are important and their experimental signature dominates the usual charge response. As a result the field distribution is modified on a distance $\sim\xi_v$ from the center of each vortex \cite{Michal}. The existence of these currents were anticipated \cite{Hou2} in the context of electrodynamics of the FFLO (Fulde-Ferrel-Larkin-Ovchinnikov) state.

The VL static linewidth is defined as
\begin{equation}
\sigma_s^{VL}=\frac{\gamma_\mu}{\sqrt{2}}[\overline{\delta h(\mathbf{r})^2}]^{1/2},\\
\end{equation}
where $\gamma_\mu=2\pi\times 135.5342\,\text{MHz/T}$ is the muon gyromagnetic ratio, $h(\mathbf{r})$ is the component of the local field parallel to the applied field $H$, overline means averaging over an unit cell of the VL, and the induction $B\simeq H$ in the limit of a large Ginzburg-Landau (GL) parameter $\kappa=\lambda/\xi$ ($\lambda$ and $\xi$ are the two length-scales of the Ginzburg-Landau theory). This quantity can be expressed through a sum involving all order Fourier components $F_{mn}$ of the field distribution in the VL,
\begin{equation}
\sigma_s^{VL}=\frac{\gamma_\mu}{\sqrt{2}}[\sum_{(m,n)\neq(0,0)}(F_{mn})^2]^{1/2}.
\label{sigmas}
\end{equation}
The components $F_{mn}$ are designated the VL form factors (FF)\cite{Michal,Yaouanc2} in the context of SANS experiments. The measurement of the first order FF $F_{10}$ at $T = 50\text{ mK}$ in \cite{Bianchi,White} has revealed a similar behavior to the one obtained in $\mu$SR experiment \cite{Spehling}: $F_{10}$ increases with field up to $4.7\,\text{T}$ to eventually decrease at the approach of the (first order) superconducting to metal transition. 

The FF calculations \cite{Michal} were performed at temperatures where the superconducting transition is second order or weakly first order $T\gtrsim1\,\text{K}$, i.e. where the GL formulation is expected to describe qualitatively the vortex lattice field distribution. In the limit $a/\xi\gg1$, with $a=\sqrt{\phi_0/B}$ the inter-vortex distance in a square VL, the FF can be written as a sum of two distinct contributions
\begin{equation}
F_{mn}=F_{mn}^{orb} + F_{mn}^{Z}.
\label{F}
\end{equation}
The first contribution is the usual charge response which gives rise to the supercurrents we called orbital. In the isolated vortex approximation and in the large-$\kappa$ limit it writes
\begin{equation}
 F_{mn}^{orb} = \frac{B\xi_v}{q_{mn}\lambda^2}K_1 (q_{mn}\xi_v),
 \label{Forb}
\end{equation}
and shows a monotoneous decrease with applied field. Here $\xi_v = \sqrt{2}\xi$ is a variational parameter that minimizes the total free energy, $q_{mn}=(2\pi/a)(m^2+n^2)^{1/2}$ for a square vortex lattice, $K_n(z)$ is the n$^{th}$ order modified Bessel function of the second kind, and $\phi_0 = 2.07\times10^{−7}\,\text{G.cm}^2$ is the flux quantum.

In contrast, the temperature and field dependent Zeeman spin contribution is
\begin{equation}
 F_{mn}^Z=\frac{\phi_0}{(2\pi\lambda)^2}\frac{\mu B}{T\ln(T_c/T)}\Im\mathfrak{m}\Psi^{(1)}\Big(\frac{1}{2}-i\frac{\mu B}{2\pi T}\Big)K_0(q_{mn}\xi_v).
 \label{FZ}
\end{equation}
It decreases with temperature, increases with field, and brings the main effect at $B\lesssim H_{c2}$ and $T$ on the order
of $T_c = 2.3\text{ K}$. Here $\mu = g\mu_B/2$ is the electron magnetic moment and  $\Psi^{(1)}(z)$ is the derivative of the digamma function. The above expression is specific to the symmetry of the order parameter and is written here for d-wave pairing. 

\begin{figure}[h!]
\centering
\includegraphics[width=8cm]{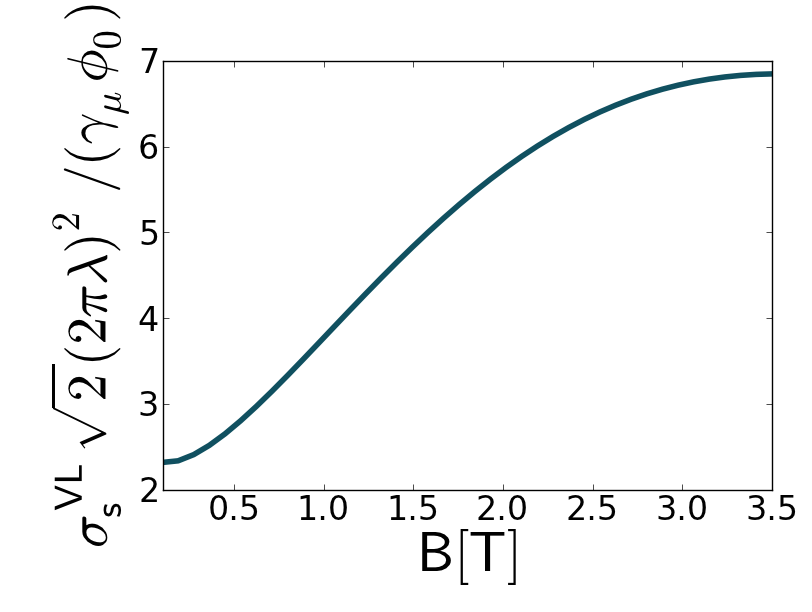}
\caption{Variation of the $\mu$SR static linewidth (Eqs. (\ref{sigmas}-\ref{FZ})). We used the temperature $T=1.3\text{ K}$, and the parameters $g=2$ and $\xi_v=50\AA$.}
\label{Plot1}
\end{figure}

The static linewidth as expressed in Eqs. (\ref{F}-\ref{FZ}) is shown in Fig. \ref{Plot1} (here for simplicity we have neglected the small field-dependence of the characteristic lengths of the Ginzburg-Landau theory).

\begin{figure}[h!]
\centering
\includegraphics[width=8cm]{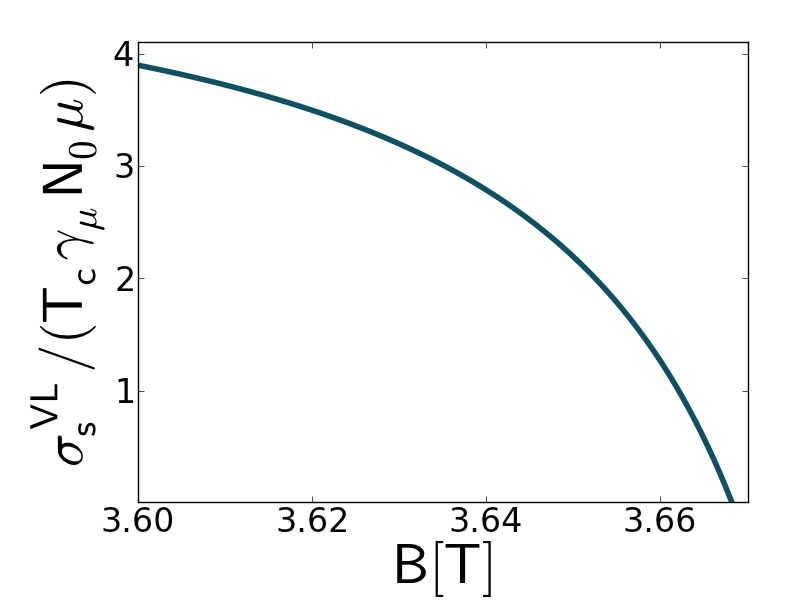}
\caption{Behavior of the $\mu$SR static linewidth close to the second-order transition line at temperature $T=1.3\text{ K}$ as obtained from Abrikosov's analysis in the Pauli limit.}
\label{Plot2}
\end{figure}

The FF given by Eqs. (\ref{sigmas}-\ref{FZ}) is found in the independent vortex approximation. It does not work near the transition to the normal state at which point $\sigma_s$ drops to zero where the transition is of the second order, or to a finite value where the transition is of the first order. In the high field limit which is close to the transition line, the main source of magnetic field inhomogeneity in the vortex lattice is the Zeeman one\cite{Hou2,Michal} 
\begin{equation}
 \delta h(\mathbf{r})=-4\pi\varepsilon(|\Delta(x,y)|^2-\overline{|\Delta(x,y)|^2}),
\end{equation}
where
\begin{equation}
 \varepsilon=\frac{N_0\mu}{2\pi T}\Im\mathfrak{m}\Psi^{(1)}\Big(\frac{1}{2}-i\frac{\mu B}{2\pi T}\Big),
\end{equation}
and $N_0$ is the non-superconducting density of states at the Fermi level.  
The Fourier decomposition of the square of the gap magnitude reads
\begin{eqnarray}
 \nonumber|\Delta(x,y)|^2&=&\overline{|\Delta(x,y)|^2}\sum_{m,n=-\infty}^{+\infty}(-1)^{m+n+mn}\\&&\times e^{-\frac{\pi}{2}(m^2+n^2)}e^{2\pi imx/a}e^{2\pi iny/a}.
\end{eqnarray}
Therefore the FFs corresponding to the Bragg peaks with indices $(m,n)\neq(0,0)$ take on the form
\begin{equation}
 \mathrm{F}_{mn}=-4\pi\varepsilon\overline{|\Delta(x,y)|^2}(-1)^{m+n+mn}e^{-\frac{\pi}{2}(m^2+n^2)},\label{FF}
\end{equation} 
and the vortex lattice static linewidth becomes simply
\begin{equation}
 \sigma_s^{VL}=\frac{4\pi s}{\sqrt{2}}\gamma_\mu\varepsilon\overline{|\Delta(x,y)|^2},
 \label{sigmasA}
\end{equation}
where
\begin{equation}
 s=\sqrt{\Big(\sum_{n=-\infty}^{+\infty}e^{-\pi n^2}\Big)^2-1}\simeq0.4247.
\end{equation}
Eq. (\ref{sigmasA}) shows explicitly that the vortex lattice contribution to the static linewidth vanishes when the transition is of the second order but shows a discontinuity when the transition is of the first order. In the former case, the gap average is known\cite{Hou2,Michal}
\begin{equation}
\overline{|\Delta(x,y)|^2}=\frac{|\alpha|}{2\beta_A\beta}.
\end{equation}
Here $\alpha$ and $\beta$ are respectively the quadratic and quartic coefficients of the Ginzburg-Landau free-energy ($\alpha$ changes sign at the transition), and $\beta_A=\overline{|\Delta(x,y)|^4}/\overline{|\Delta(x,y)|^2}^2$ is the Abrikosov parameter. It is $\beta_A^\square=1.18$ for a square vortex lattice and $\beta_A^\triangle=1.16$ for a triangular lattice. Then it follows
\begin{equation}
 \sigma_s^{VL}=\frac{2\pi s\gamma_\mu}{\sqrt{2}\beta_A}\frac{|\alpha|\varepsilon}{\beta},
\end{equation}
which is shown in Fig. \ref{Plot2}.

While experiment was made at $T = 20\text{ mK}$, our results are shown at $T = 1.3\text{ K}$ where the GL formulation can be applied. Our model yields an effect that is observable in this regime and it is expected to be enhanced at lower temperatures. It would be interesting to
obtain experimental information on the static linewidth variations at higher temperatures.

As a final remark, the authors in [Ref. 1] assumed a temperature-independent static linewidth to extract the temperature dependence of the relaxation rate dynamic contribution. However it is known from the temperature dependence of $\lambda$ [Ref. \cite{Abrikosov}] that $\sigma_s$ varies as $1−[2\pi\Delta(0)/T]^{1/2}\exp[−\Delta(0)/T]$ for $\mu B\ll T$ and $T\ll T_c$ ($\Delta(0)$ is the value of the superconducting gap at $T = 0$), and as $(T_c−T)/T_c$ for $\mu B\ll T$ and $T_c − T\ll T_c$. For higher fields, we obtain from Eqs. (\ref{sigmas}-\ref{FZ}) significant and qualitatively similar variations.

\emph{Acknowledgment}. We are grateful to V. Mineev for careful reading of the manuscript.

\end{document}